\documentclass[twocolumn,showpacs,prl]{revtex4}

\usepackage{graphicx}
\usepackage{dcolumn}
\usepackage{amsmath}
\usepackage{latexsym}

\begin{document}

\title{Polydispersity-linked Memory Effects in a Magnetic Nanoparticle System}

\author{S.Chakraverty$^{1}$, A.Frydman$^{2}$, M.Bandyopadhyay$^{1}$, S.Dattagupta$^{1}$, Surajit Sengupta$^{1}$ and P. A. Sreeram$^{1}$\\
}
\affiliation{
1.Nanoscience Unit, S. N. Bose National Centre for Basic Sciences, \\
Block-JD, Sector-III, Salt Lake,
Kolkata - 700098,India.\\
2. Dept.of Physics, Bar Ilan University, Ramat Gan 52900, Israel
}

\date{\today}

\begin{abstract}
\noindent We have performed a series of measurements on
the low temperature behavior of a magnetic
nano-particle system. Our results show striking memory effects in
the dc magnetization. Dipolar interactions among the nano-particles 
{\em suppress} the memory effect. We
explain this phenomenon by the superposition of different
super paramagnetic relaxation times of single domain magnetic nano- particles.
Moreover, we observe a crossover in the temperature 
dependence of coercivity. We show that a dilute dispersion of 
particles with a flat size distribution yields the best memory. 
\end{abstract}

\maketitle

Single domain magnetic nano-particles\cite{nano,nano1,nano2,nano3,nano4,nano5,nano6,nano7} possess relaxation 
times that depend exponentially on the volume\cite{relax1,relax2,relax3,sdg}. Thus polydispersity leads to a distribution of relaxation times\cite{sun,sasaki}, those larger than the measurement time yielding 'frozen' behavior, and those shorter
giving rise to 'magnetic viscosity'\cite{relax3,sdg}. A given sample then
displays strong memory effects, which are reported here. Our results are based on the measurements of temperature-dependent magnetization during
cooling and heating cycles. 
These memory effects may have important device applications\cite{nano2} in the future.
\vskip .01cm

\noindent
In this Letter we report results on magnetization and coercivity measurements 
in systems of $NiFe_{2}O_{4}$
particles (mean size$\simeq$ 3nm.) embedded in a $SiO_{2}$ matrix. 
Both measurements show strong history dependent effects depending on the separation between the particles and hence their mutual interaction.
 We  prepare the samples by using a sol-gel\cite{sol-gel} technique. The ratio of $NiFe_{2}O_{4}$ to $SiO_{2}$ is 1:1 and 3:7 by weight, for  sample A and sample B respectively. The ferrite to marix ratio is  altered in order to tune the interparticle interaction. The phase of the samples is identified by X-ray diffraction\cite{xray} using a Philips PW 1710 diffractometer with Cu K{$_\alpha$} radiation ({$\lambda$} = 1.54 A$^o$). The average particle size is estimated by X-ray diffraction as well as by a JEM-200-CX  transmission electron micrograph(TEM). DC magnetization measurements are performed on a Quantum Design superconducting interference device magnetometer (MPMS) from 300k to 4k. \\
The X-ray diffraction spectrum confirms that our sample is indeed in a single-phase of ferrite NiFe$_2$O$_4$ with no residual $\alpha$-Fe$_2$O$_3$. The average particle size of the nanophase NiFe$_2$O$_4$, estimated from the broadening of X-ray diffraction peak is $\sim$ 3nm for both the samples. The TEM micrograph of the samples suggests that the particles are spherical in shape and the particle size follows a log-normal\cite{sasaki} distribution. The interparticle separation measured from TEM micrographs are $\sim$ 5nm for sample A and $\sim$ 15nm for sample B.\\
\noindent
The magnetization measurements are carried out in accordance with the
following cooling and heating protocol. At T=300K
($T=T_{\infty}$), a small magnetic field ($H = 50$~Oe) is applied and the
magnetization(M) measured. Keeping the field on, the temperature(T) is lowered 
continuously at a steady rate to $T_{n}$ and M is simultaneously measured upto temperature $T_{n}$.
Thus one obtains M versus T in the cooling regime
$(T_{n}\le T\le T_{\infty})$. At T$_{n}$ the field is switched
off and the drop of M is monitored for several ($\approx 4$) hours. 
Subsequently, the magnetic field is {\em switched} back on and
M(T) versus T is mapped in the cooling regime($T_{n-1}\le T\le
T_{n}$). At T$_{n-1}$ the field is switched off again and the
process of measurement repeated, until the lowest temperature
T$_{0}$ is reached. Thus, one obtains field-cooled response and zero-field
relaxation of the magnetization as a function of temperature.
At the end of the cooling cycle, at T$_{0}$, the field is turned on and
M(T) monitored as the system is heated from T$_{0}$ through
T$_{n-2}$,T$_{n-1}$,T$_{n}$ and eventually to T$_{\infty}$, the magnetic
field remaining on throughout.
Our results are shown in Fig.1, for two distinct
values of  interparticle spacing  5nm.(sample A), and 15nm (sample B). The heating path
surprisingly shows wiggles in M(T) at all the T steps
T$_{n-2}$,T$_{n-1}$,T$_{n}$ where H was earlier switched off during
cooling, apparently retaining a memory of the temperature steps at
which the cooling was arrested.
\begin{figure}[t]
\begin{center}
\includegraphics[width=7.0cm,angle=270]{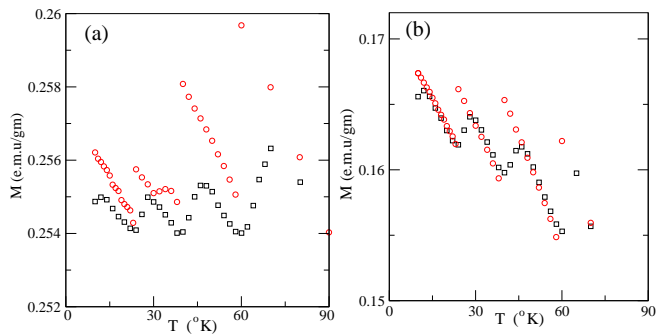}
\end{center}
\vskip -.6cm
\caption{(color on-line) Experimental  M(T) curves during cooling ($\Box$ red) 
in a small magnetic field $H = 50$ Oe and zero-field heating ($\Diamond$ black) 
for the (a) interacting and (b) non-interacting 
samples showing prominent memory effects. A constant heating/cooling rate of 
$2^{\circ}$ K was maintained except at $60^{\circ}$, $40^{\circ}$ and 
$20^{\circ}$ K where the cooling was arrested for $4$~h duration at each 
 temperature during which time $H$ was switched off. }
\label{memoryeffectexpt}
\end{figure}

\noindent
One tantalizing aspect of our results is that memory effects are
more prominent for sample B than for sample A, although in the latter the average inter-particle distance is smaller and hence the dipolar interaction larger.
Recently Sun {\em et. al.}\cite{sun} have reported very similar history
dependent effects in the magnetization measurements of
a monolayer of sputtered permalloy($Ni_{81}Fe_{19}$) clusters on a
$SiO_{2}$ substrate. These authors attribute the disparate cooling
and heating histories to aging and concomitant memory effects
found in a spin glass phase\cite{spngls1}. Spin glass transitions are known to
occur due to disorder and frustration\cite{spngls2} in dilute magnetic alloys
that are characterized by a complicated free energy landscape with
deep valleys and barriers. Strongly nonequilibrium memory
dependent behavior ensues as a result of the system getting trapped in a 
deep valley\cite{spngls3} such that the relaxation time($\tau$)
for deactivation becomes long compared to experimental time scales of 
measurement.
\vskip .01cm

\noindent
Our interpretation of the results shown in Fig.\ref {memoryeffectexpt} is very different from that of \cite{sun} . We demonstrate below that the observed
phenomena (by us as well as by other groups) are {\em not} connected to complicated spin glass type interactions but can be simply attributed to a superposition of relaxation times, arising from particle size distribution, as it were in {\em noninteracting} single-domain magnetic particles. Experimentally it is known\cite{sasaki} that nano particle sizes are usually distributed according to a log-normal distribution. However, we 
show below that the exact form of the distribution is irrelevant for explaining
the memory effect. In fact, in order to keep the analysis simple and to obtain
a clear understanding of the physics it is sufficient to take a simple size
distribution consisting of two delta function peaks so that there are only
two kinds of particles ``large'' (volume V$_1$) and ``small'' (V$_2$).
Correspondingly we have only two relaxation times
$\tau = \tau_{1}$ and $\tau_{2}$ in our model, if we remember\cite{relax1,relax2,relax3} that:
\begin{equation} 
\tau(V)\propto \exp[\frac{(KV\pm \mu HV)}{K_{B}T},
\end {equation} 
where K is the anisotropy energy, $\mu$ is the
magnetic moment per unit volume, H is the applied magnetic field
and $K_{B}$ is the Boltzmann constant\cite{sdg}.
\begin{figure}[h]
\begin{center}
\includegraphics[width=7.5cm,angle=270]{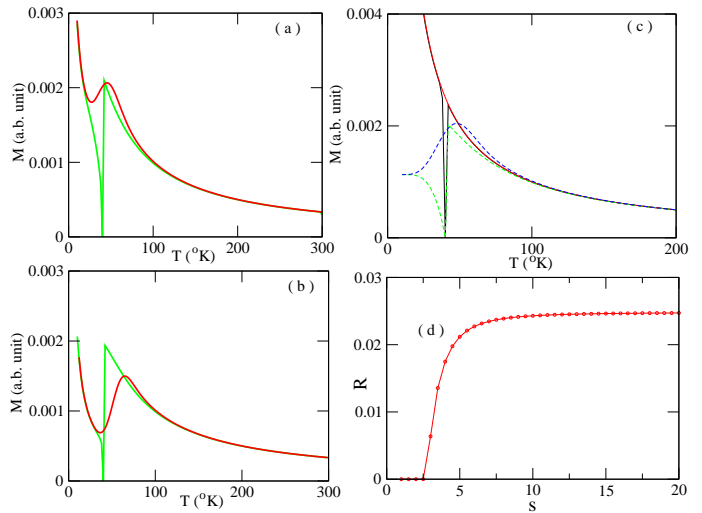}
\end{center}
\vskip -.6cm
\caption{(color on-line) Simulated  M(arbitrary unit) vs. T curves during cooling  and heating
 for the (a) interacting and (b) non-interacting cases:  
 The curve (c) shows the 
various contributions to the total magnetization (thick solid line)of interacting sample(A) coming 
from the fast (thin solid line) and slow (dashed line) particles. 
The theoretical curves (a) - (c) have been calculated using a double delta 
function distribution of particle sizes. Curve 
(d) shows a plot of the recovery parameter $R$ (see text) as a function of
the width(s) of a Gaussian particle size distribution.}
\label{memoryeffectsimul}
\end{figure}
 Our interpretation of the observed results hinges on
the premise that the time $\tau_{1}$ is much larger
than the measurement time while $\tau_{2}$ is much smaller, at the 
lowest measured temperature (T$_{0}$).
Both $\tau_{1}$ and $\tau_{2}$ are expected to be smaller than the
measurement time at the highest temperature T$_{\infty}$.
Therefore, in the intermediate temperature domain ($T_{0}\le T\le
T_{\infty}$), the small particles  equilibrate
rapidly, thus showing superparamagnetic viscosity\cite{sdg} while the large
particles are 'blocked'. This is observed in Fig.\ref {memoryeffectsimul}(c) where we have plotted computer simulations\cite{s.c.} of M(T) separately for the two sets of  interacting(separation distance of $\simeq$5nm) particles under the same cooling and heating regimens. Here we choose the 
temperature $T^*$ at which $H$ is switched off such that the blocking 
temperatures\cite{relax1,relax2} corresponding to the two different particle sizes flank $T^*$.The simulations are based on standard rate theory calculation for the time dependent magnetization\cite{s.c.}. When H is
zero, both sets of particles relax to M=0. However, when H is
turned on, particles 1 are blocked(M=0) while 2 show facile
response. As T is increased again, M for particles 2 decreases
with T while M for particles 1 initially increases before dropping
off. The resultant graph is a superposition (see Fig.\ref {memoryeffectsimul} (c)) of a monotonically decreasing curve and a hump, thus producing a wiggle. This effect is seen only when the temperature of arrest is in-between the two 
respective blocking temperatures, in conformity with the findings of \cite{sasaki}. \\
We have performed measurements on the same
system but now with increased interparticle separation
($\simeq$15nm.)(see Fig.\ref {memoryeffectexpt}(b)),the simulation results of which are shown in Fig.\ref {memoryeffectsimul}(b).
\vskip .01cm

\noindent
The resultant interaction effect due to dipole-dipole coupling, not considered
in \cite{sasaki}, is also quite distinct from the quenched-in
disorder mediated interactions proposed in \cite{sun}. Since the
dipole moment of a single-domain particle is proportional to its
volume, the effect of interaction, within a mean-field picture, may be 
incorporated by adding a term proportional to $V^{2}$ in the
exponent of $\tau(V)$.
Thus, even small particles ($V_{2}$) can
now have $\tau_{2}$ {\em larger} than the
measurement time. This becomes more prominent at lower
temperatures. Therefore, the blocking temperatures for both
particles 1 and 2 are now shifted to higher T, thereby causing the
wiggles to disappear. This is consistent with the results
of Fig.\ref {memoryeffectexpt} which show that the memory effects are {\em stronger} for the non-interacting particles. We conclude then that the unexpected
wiggles seen in the cooling and heating
cycles of M(T) versus T have much less to do with interaction effects but
more to do with polydispersity of the sample.
\vskip .01cm

\noindent
How crucially does the nature of the particle size distribution function
affect the magnetization recovery during the zero field heating cycle ? In
order to answer this question we first quantify the memory effect by 
defining a parameter,
\begin{equation}
R = \Theta(\frac{dM}{dT}|_{T=T_n})\frac{dM}{dT},
\end{equation}
where $\Theta(x)$ is the Heaviside step function. The parameter $R$ 
measures the positive slope of the M(T) curve during zero field heating.  
We have calculated $R$ using a Gaussian size distribution centered at 
$V = V_0$ and with width $s$. Our results for $R$ are shown in 
Fig.\ref {memoryeffectsimul}(d) for a particular choice of $V_0$ as a function of $s$. We observe that $R$ increases with the width of the distribution
and saturates quickly. In this regime, $R$ is almost independent of $V_0$ and accordingly, the detailed nature of the distribution.
We conclude that the memory effects will be best seen in samples with a 
dilute dispersion of particles but a very wide (flat) distribution of
sizes. Indeed in this limit the relaxation is known\cite{sdg} to be prominently dominated by magnetic viscosity characterized by a logarithmic relaxation in time. Not surprisingly, a logarithmic relaxation has been observed
in the experiments of Sun {\em et. al.}\cite{sun} although the interpretation
offered is different from ours. 
\vskip .01cm

\noindent 
In order to substantiate our interpretation of the  M vs. T data
we have carried out hysteresis measurements and thereby coercivity
estimation for both the interacting sample A and non-interacting
sample B.
\begin{figure}[t]
\begin{center}

\includegraphics[width=6.5cm,angle=270]{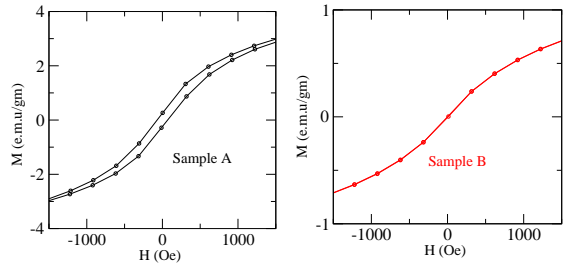}
\end{center}
\vskip -.6cm
\caption{(color on-line) Room temperature M-H curve of samples A and B. The
hysteretic response of sample A points to the presence of strong dipolar 
interactions.}
\label{mhloop}

\end{figure}
The {\em room-temperature} DC magnetizations versus the applied magnetic field
for both samples are   shown in Fig.\ref {mhloop}.
Clearly, for sample B the relaxation times 
are shorter than the measurement time, at 300K. Thus,
there is no hysteresis loop and the coercivity
(measured by the width along the abscissa on the
zero-magnetization line) is also zero. On the other
hand, for sample A, we observe a
non-zero coercivity even at 300K 
due to the slowing down of relaxation because
of the presence of an additional term proportional to $V^{2}$
in the exponent of $\tau(V)$ as mentioned above.\\
\begin{figure}[t]
\begin{center}
\includegraphics[width=6.0cm,angle=270]{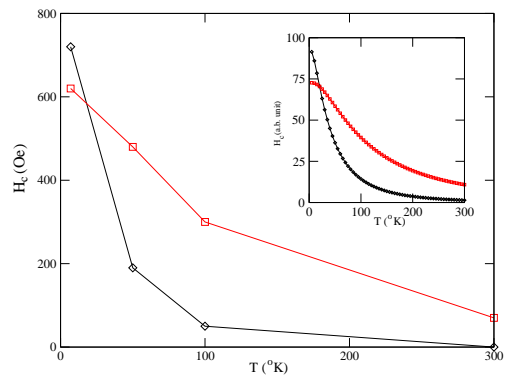}
\end{center}
\vskip -.6cm
\caption{(color on-line) Coercivity($H_{c}$) as a function of temperature
for the interacting ($\Box$) and non-interacting ($\Diamond$) samples.
The corresponding curves ($H_{c}$ in arbitrary unit) obtained from our theory assuming a double delta function particle size distribution is shown in the inset.}
\label{coercivity}
\end{figure}

\noindent
Next we repeat the above measurements down
to 4K, using a SQUID magnetometer. The coercivity($H_{c}$) is plotted as
a function of temperature(T),  in Fig.\ref {coercivity}.
 Because relaxation slows down for both sample A and sample B, $H_{c}$ increases with decrease of T (Fig.\ref {coercivity}). 
The coercivity of the interacting
sample A is larger than that non-interacting sample B for
temperatures greater than 25K. However, at T=25K a surprising
{\em crossover} is detected, where the coercivity for
sample B shoots above that for sample A. We suggest
that the reason for this behavior is that the
term H in the exponent of $\tau(V)$ is replaced by
H+$\delta$H, where the mean field $\delta$H, arises from
interaction:
\begin{equation}
\delta H=\mu V^{2}\gamma \tanh(\frac {\mu V(H+\delta H)}{K_{B}T})
\end{equation}
where $\gamma$ is a parameter that depends on the mean separation between the particles. 
The tanh term augments the V$^{2}$ term
below 25K, making the larger particles relax so slowly that they
don't respond to H at all. Therefore, the larger
particles are 'frozen out' from further consideration,
making the mean
relaxation time in the interacting case even smaller than that for
the non-interacting case.  This somewhat nonintuitive conclusion
is further confirmed by
our simulated coercivity computation, shown in Fig.\ref {coercivity} (inset).\\
To verify our argument further we perform
a separate set of experiments on both  samples A and B as follows. We field-cool the samples down to 10K from 300K in the presence of $H =  100$~Oe. At 10K
the magnetic field is switched off and the
relaxation of the magnetization measured. We
find that the average relaxation time obtained by forcing an exponential fit to our data of sample A is 100 min and
that of sample B is 25min. We then heat the samples to
300K, and  cool it back down to 10K at zero magnetic field.  At
10K we switch on the magnetic field and wait for
2h. The magnetic field is then switched off and
the magnetization measured. The relaxation time of sample
B remains 25min. but the relaxation time of sample A decreases to 30
min. This result is consistent with the reasoning
described in the above paragraph. Therefore, for the
low-temperature interacting system, larger particles are rendered magnetically inactive.\\
In conclusion, the strong history dependent effects seen in
magnetization and coercivity measurements in $NiFe_{2}O_{4}$
magnetic nanoparticles are interpreted as being  due to arrested 
Neel relaxation. Our model is
dramatically simplified by choosing just two volumes of the
particles, on either side of the 'blocking' limit. Further
corroboration of the proposed mechanism is achieved by performing
measurements on an interacting system. Our results suggest that
either by tuning the interaction (through changing inter-particle
distance) or by tailoring the particle size distribution, these
nanosized magnetic systems can be put to important application in
memory devices. In particular, a flat volume
distribution can be of great utility than a monodispersed distribution with a 
single sharp peak.\\
SC thanks Kalyan Mandal for guidance on sample preparation. We  acknowledge support from the Department of Science and Technology through its "Nanoscience Initiatives". SS acknowledges financial support from DST grant no: SP/S2/M-20/2001


\begin{thebibliography}{99}
\bibitem {nano} A.N.Goldstein,{\em Hand Book of Nanophase Materials} (New York:Marcel Dekker Inc.)(1997).
\bibitem {nano1} {\em Magnetic Properties of Fine Particles}, edited by J.L.Dormann and D.Fiorani (North-Holland, Amsterdam, 1992).
\bibitem {nano2}K.M.Unruh and C.L.Chein, {\em Nanomaterials: Synthesis, Properties and Applications}, eds., A.S.Edelstein, R.C.Cammarata(Bristol: Institute of Physics).
\bibitem {nano3}I.S.Jacobs and C.P.Bean, in {\em Magnetism III (eds.) G.T.Rado and H.Suhl} (New York: Academic)(1963).
\bibitem {nano4}{\em Physical Principles of Magnetism}, A.H.Morrish, John Wiley,New York,1965.
\bibitem {nano5} J.Frankel and J.Dorfman, Nature(London){\bf 126}, 274(1930).
\bibitem {nano6} C.Kittel, Phys.Rev.{\bf 70}, 965(1946).
\bibitem {nano7} C.P.Bean and J.D.Livingstone, J.Appl.Phys.{\bf 30}, 120S(1959).
\bibitem {relax1} Steen Morup, Elisabeth Tronc, Phys. Rev. Lett. {\bf 72},3278(1994).
\bibitem {relax2} L.Neel, Ann.Geophys.{\bf 5}, 99(1949); Adv.Phys.{\bf 4}, 191(1955)
\bibitem {relax3} E.P.Wohlfarth, J.Phys.{\bf F10},L241(1980).
\bibitem {sdg}R. Street and J. C. Woolley, Proc. Phys. Soc London, Sec A {\bf 62}, 562 1949. Also reviewed in S.Dattagupta,{\em Relaxation Phenomena in Condensed Matter Physics, Chapter XV}, Academic Press, Orlando(1987).
\bibitem {sun} Y.Sun, M.B.Salamon, K.Garnier and R.S.Averback, 
Phys. Rev. Lett. {\bf 91},167206(2003).
\bibitem {sasaki}M.Sasaki, P.E.Jonsson, H.Takayama and P.Nordblad, arXiv:cond-mat/0311264 v2.
\bibitem {sol-gel}J.D.Wright and J.M.Sommerdijk, {\em Sol-Gel Materials,Chemistry and Applications}(London:Taylor and Francis)(2001).
\bibitem {xray}B.D.Cullity, {\em Elements of X-ray Diffraction}(1978)
\bibitem {spngls1}A.P.Young, {\em Spin Glasses and Random Fields}, eds., World Scientific, Singapore (1987).
\bibitem {spngls2} K.H.Fischer and J.A.Hertz,{\em Spin Glasses},Cambridge University Press (1991).
\bibitem {spngls3}See for instance, K.Binder and W.Kinzel in Heidelberg Colloquium on Spin Glasses (J.L.Van Hemmen and I.Morgenstern),eds.,Springer-Verlag,Berlin and New York,p279 and references therein.
\bibitem {s.c.} S. Chakraverty, S. Chatterjee, S. Dattagupta, A. Frydman, (Submitted)
\end{thebibliography}
\end{document}